\documentstyle[12pt,aaspp4,epsf]{article}

\def\cite{{\it cite}}

\def\kms{km s$^{-1}$}

\def\gtsim{\ {\raise-0.5ex\hbox{$\buildrel>\over\sim$}}\ }
\def\ltsim{\ {\raise-0.5ex\hbox{$\buildrel<\over\sim$}}\ }
\slugcomment{Submitted to The Astrophysical Journal Letters}

\begin{document}

\title{Detection of Lithium in a Main Sequence Bulge Star\\
Using Keck~I as a 15m Diameter Telescope}

\author {
D. Minniti$^{1}$,
T. Vandehei$^{2}$,
K.H. Cook$^{1}$, 
K. Griest$^{2}$, 
C.  Alcock$^{1}$
}

\altaffiltext{1}{Lawrence Livermore National Laboratory, Livermore, CA 94550\\
E-mail:  alcock, kcook, dminniti@llnl.gov}

\altaffiltext{2}{Department of Physics, University of California, San Diego, CA 92093 \\
E-mail: griest, vandehei@astrophys.ucsd.edu}

\begin{abstract}
The bulge contains the most chemically evolved old 
stellar population in the Milky Way.
Thanks to microlensing, it is now possible to obtain high resolution 
echelle spectra of
bulge stars near the main sequence turn-off, and study the abundance
of elements that are affected by stellar evolution, such as lithium.
We have observed with the HIRES spectrograph on the Keck~I 10m telescope 
the source star of the MACHO microlensing event 97BLG45 while 
it was magnified by $\sim 1$ magnitude.  Here we report the detection of
 the Li~I line at $\lambda 6707.8$ \AA\ in the echelle spectrum of
this star, and measure the
bulge lithium abundance for the first time: 
$A(Li) = 2.25 \pm 0.25$.
\end{abstract}

\section{Introduction}

Sneden et al. (1995) summarized the new opportunities for stellar population
studies with high spectral resolution at very large telescopes. The
advent of microlensing helps push these opportunities even further.

The greatest advantage of microlensing is the achromatic magnification of
the brightness of the source, allowing the observation of objects that would 
otherwise be out of reach.
The net effect of adding such a lens in front of the telescope is to
increase the diameter of this telescope. 
Here we present a specific application of this concept: we have obtained 
a high resolution spectrum of a magnified ($A\approx 2.7$)
bulge main-sequence star at a distance of $\sim 8$ kpc.
This is effectively using the Keck~I telescope as a 15m diameter telescope.
There is simply no other way to obtain such
high resolution spectra of bulge main sequence stars.

While bulge red giants are within 
reach of the echelle spectrographs of 4m to 10m class
telescopes, the bulge main sequence stars demand telescopes with 15m diameter
or larger. It is, however, interesting to observe bulge main sequence 
stars rather than giants for two main reasons.
First, for stars near the main sequence turn off 
it is possible to estimate ages based on accurate photometry and chemical 
abundances. Unfortunately, the ages of red giants cannot be determined.
Second, the atmospheric abundances of some elements are affected by stellar
evolution.  One such element is lithium, which is depleted due to the
onset of convection as the stars evolve up the giant branch
(e.g. Spite \& Spite 1982, Hobbs \& Pilachowski 1988, Boesgaard 1991).

The chemical composition of dwarf stars in the halo and disk components of
the Milky Way has been extensively studied because these stars are represented
in the Solar neighborhood (Wheeler, Sneden \& Truran 1989,
Edvarsson et al. 1993). Dwarfs in the bulge have remained beyond
reach of existing telescopes because they are intrinsically too faint.
The bulge, however, is of particular interest, having high metal
(McWilliam \& Rich 1994) and He abundance (Minniti 1995). Being at the bottom 
of the Galactic potential well, it is the place where one expects to find the 
most chemically evolved populations, formed from gas leftover after the
formation of the halo (Minniti 1996). It is our long term goal 
to take advantage of microlensing in order to
assemble a bulge sample like the unevolved F and G disk stars 
of Edvarsson et al. (1993),
and define the chemical composition and enrichment history of the bulge.
Such a sample would be invaluable to study the Galactic evolution 
of lithium (Hobbs \& Pilachowski 1988, Lambert et al. 1991, Boesgaard 1991,
Balachandran 1995, Spite 1995).

Microlensing will allow for the first time the
determination of detailed chemical compositions for
bulge main sequence stars {\it in situ}. 
Note that this is possible because
microlensing is achromatic, and does not affect the stellar 
spectrum (Benetti et al. 1995)
One particular aspect of this program is that it does not need to be a
target of opportunity. The MACHO project is producing enough 
microlensing alerts during the
bulge season, that at any given night there are several magnified candidates. 
However, a disadvantage of microlensing is that the observations are 
non-repeatable, because the same star is not magnified twice.

\section{The Source Star 97BLG45}

The star 97BLG45 was alerted on by the MACHO collaboration in late July 
1997.  Data on the alert was taken from the MACHO Project homepage at
{\bf http://wwwmacho.mcmaster.ca}. The position, photometry, and
other relevant information are listed in Table 1.
The baseline magnitude of this star is listed, although
the star was 1 mag brighter at the time of the observations.

The star 97BLG45 is located in the overlap region between MACHO fields 136 and 
142, lying at a projected distance of $1.08$ kpc from the Galactic center.
The position in the MACHO color-magnitude diagram of the bulge (see 
Alcock et al. 1997a), indicates that this star is likely to be located 
near the bulge main sequence turn-off. It is too blue to
be a disk main sequence star. If it were a disk main sequence star, it would 
be very distant, and more than one kpc above the Galactic plane, which is
unlikely. Also, the location in the color-magnitude diagram is not consistent 
with this star belonging to the Sgr dwarf.

The reddening in MACHO field 142 has not been measured yet. 
We estimate the reddening in this field to be $E(V-R)=0.20$
by extrapolating the reddening measurements of Alcock et al.
(1997b). These reddening determinations are based on the mean colors of RR Lyrae
in bulge fields.  The adopted reddening value implies that 97BLG45 has
an absolute magnitude $M_V = +4.4$ if it is located at a distance of
about 8 kpc. This places it near the bulge main sequence turn-off.

The observations of 97BLG45 were taken on the first half of the night
of August 18, 1997, under
sub-arcsecond seeing conditions. We used the Keck~I 10m telescope with the
HIRES echelle spectrograph (Vogt et al. 1994).
The Li~I line at $\lambda 6707.8$ \AA\ is located in the echelle order \#22,
with a dispersion of $0.094$ \AA\ $pix^{-1}$, 
yielding a resolution of $0.24$ \AA, as measured by the FWHM of
the spectral lines. The final spectrum consists of the
combination of 6 half hour spectra, taken with airmasses ranging from 1.5 to 3. 
Some of the spectroscopic parameters are listed in Table 2,
and a detailed description of the data reduction will be published elsewhere
(Minniti et al. 1998). 
The final combined spectrum of 97BLG45 has resolving power $R=27000$, and
$S/N = 50$ per resolution element.
We observed IAU  radial velocity standards and abundance standards from the
list of Edvarsson et al. (1993). At twilight 
we also obtained a spectrum of the Sun for comparison, and of
rapidly rotating stars. The latter are used for identifying and removing the
atmospheric absorption lines. The telluric emission lines are accounted
for by the sky present in the columns adjacent to the object for every order.
There are no blemishes in the flats, or sky lines next to the Li~I 6707.8 line 
that would affect our measurement of this line.
The background sky and scattered light amount to 24 counts  per pixel 
in the mean in this region of the spectrum, or less than $10$\% of the total
counts in this spectrum.

\section{The Spectrum of 97BLG45}

Figure 1 shows the final combined spectrum of a fraction of the echelle 
order \#22, with the Li~I and neighboring Fe~I lines indicated. 
The Li~I line shown in Figure 1 is real, judging by the presence of Fe~I 
lines with comparable equivalent width in this region of the spectrum (Table 2).
In particular, the equivalent width of the Fe~I line at $\lambda 6705$ \AA\
is similar to that of the Li~I line at $\lambda 6707.8$,
justifying our claim of detection of this line.

We measure an accurate radial velocity $V=89$ \kms\ for 97BLG45 by centroiding 
several strong lines in this region of the spectrum.  This radial velocity is
consistent with the observed kinematics of the Galactic bulge (Minniti 1996).
The observed velocity is too large for a normal disk star, favoring bulge 
membership, and ruling out membership in the Sgr dwarf ($V=160$ \kms).

The star 97BLG45 is single-lined, and has negligible rotation.
The rotation was estimated by comparing the FWHM of the strong
lines with the emission and absorption lines of the telluric sky spectrum,
and with the FWHM of the  ThAr comparison lamp lines.

The temperature of 97BLG45 is estimated in different ways: 
(1) Using the optical photometry, $(V-R)_0 = 0.40$. From the recent calibrations of
Clementini et al. (1995), we derive $T_{eff} = 5800$ K. 
(2) Using the equivalent width
of $H_{\alpha}$ and $H_{\beta}$. This is rather uncertain, because
of the difficulties of determining the true continuum for such broad lines.
(3) Comparing the line profiles of $H_{\alpha}$ and $H_{\beta}$
with those of the Solar spectrum. This direct comparison reveals that $T_{eff} >
T_{\odot}$.

We adopt 
$T_{eff}= 6000$K, with an 
estimated conservative error of $\sigma_{Teff} = 150$ K.
The complete spectral analysis will yield another independent value of
the stellar temperature (Minniti et al. 1998).
The $T_{eff}$ measured for 97BLG45 places it in the Spite plateau 
(Spite \& Spite 1982). Using this temperature,
we are able to measure the star lithium abundance, as detailed below. 

\section{The Lithium Abundance of 97BLG45}

The present resolution does not allow the separation of the blend of
the Li~I resonance doublet at $\lambda 6707.76$ \AA\ and $\lambda 6707.91$ 
\AA\ with the Fe~I line at $\lambda 6707.41$ \AA\ (King et al. 1997).
We measure the equivalent width of the Li+Fe blend at $\lambda 6707$ \AA\ to be 
$W_{6707}=58$ m\AA.
The error in this measurement is dominated by the choice of the continuum,
and we conservatively adopt $\sigma_{W} = 15$ m\AA .
The contamination by the Fe~I line at $\lambda 6707.4$ \AA\ is taken into
account following the procedure of Soderblom et al. (1993). We find
$W_{Fe6707.4}=11$ m\AA, yielding $W_{Li6707.8}=47\pm 15$ m\AA.

The lithium abundance in 97BLG45 is measured using the tables of Soderblom et 
al.  (1993) and Ryan et al. (1996). We find $A(Li)=12+logN(Li)/N(H)=2.25$,
value that is plotted as a large star in Figure 2.

The errors in the lithium abundance are dominated by the errors in effective
temperature and in equivalent width. The error $\sigma_{W} = 15$
m\AA\ contributes $\sigma_{Li} = 0.2$ dex. The error $\sigma_{Teff} = 150$ K
contributes $\sigma_{Li} = 0.15$ dex. We then conclude that $\sigma_{Li} = 0.25$
dex. This error is relatively large compared with other studies because the
target star is extremely faint. The detection, however, is comparable to 
that of Pasquini et al. (1997) in 47~Tuc.

For comparison, Figure 2 also shows the lithium abundances $vs$ temperature 
plane for stars in  the Hyades taken from Balachandran (1995),
47Tuc from Pasquini \& Molaro (1997), and
M67 from Pasquini et al. (1997), along with the Sun.
The position of 97BLG45 in Figure 2 is well below the sequence of young
open clusters such as the Hyades (with $700$ Myr), but it is consistent with 
older populations such as the open cluster M67 (with $4.7$ Gyr), and
the old globular cluster 47~Tuc (with $13$ Gyr).

The Li~I line is stronger than the one in the 
Solar spectrum. Since the temperatures of these two stars are not very
different, the lithium abundance of 97BLG45 is well above the Solar
value, as shown in Figure 2. In fact, if the Li I line were as weak as in
the Sun, we would not have detected it.

The most recent determination of the primordial
lithium abundance plateau from Bonifacio \& Molaro (1997),
$A(Li)=2.238\pm 0.06$, is also plotted in 
Figure 2 (see also Ryan et al. 1996).  The lithium abundance of 97BLG45 
is consistent with the mean primordial lithium abundance given
by these authors within the errorbars.

While these comparisons are interesting, based on a single measurement we
cannot draw any
conclusions about the lithium dependence with age and chemical composition
in the bulge. 
The lithium dispersion in the halo stars --Pop II-- is very small in 
comparison with that of the disk stars --Pop I-- (Spite \& Spite 1982,
Lambert et al. 1991, Spite 1995).
Clearly, a sample of about 10 objects like 97BLG45 is
needed to accurately define the Spite plateau and measure the lithium
dispersion in the bulge population. This is currently within reach of
the Keck~I telescope: for stars magnified by microlensing it would be possible
to study the lithium-metallicity dependence 
(e.g. Lambert et al. 1991), and the lithium-age dependence (Boesgaard 1991)
in comparison with the other Galactic components.
Such a sample would also allow the measurement of the abundances of different
elements in the bulge following the precepts of Edvarsson et al. (1993).
Most importantly, it may then be possible to establish if the bulge formed 
from gas previously enriched in the halo. 

\section{Conclusions and Future Prospects}

We have introduced a specific application of
microlensing as a tool for the study of faint objects, obtaining 
the first determination of the lithium abundance in the Milky Way bulge: 
$A(Li) = 2.25 \pm 0.25$. 
This measurement is based on an echelle spectrum of the microlensing 
event 97BLG45 of a star near the bulge main sequence turn-off.
We have achieved the faintest limit possible with current equipments, 
using Keck I, the largest existing telescope in the world, with the longest 
allowed exposure time in the half night, and the bulge main sequence star 
with highest magnification at the time.

This opens up new possibilities for the study of the chemical composition and
enrichment history of the Milky Way bulge, to complement similar
studies of the other major galactic components: halo (Wheeler et al. 1989),
and disk (Edvarsson et al. 1993).
Furthermore, it would not be unreasonable to expect to obtain echelle spectra
and measure element abundances in individual stars in nearby galaxies 
that are magnified by microlensing. Even individual
red giants in the M31 bulge, or subgiants in the Magellanic Clouds
that are magnified by factors of $> 10$ would be within reach.

\acknowledgements{
This work would not have been possible without the MACHO microlensing alerts,
mantained by A. Becker and the MACHO team.
We would like to thank S. Burles and S. Marshall for help with the HIRES
software.
Thanks also 
to the W.M.Keck Observatory where the observations
were obtained.  Work at LLNL is supported by DOE contract W7405-ENG-48,
and work at UCSD is supported by DoE grant DE-FG03-90ER40546
and by the Alfred P. Sloan Foundation. 
}

\begin{figure}
\plotone{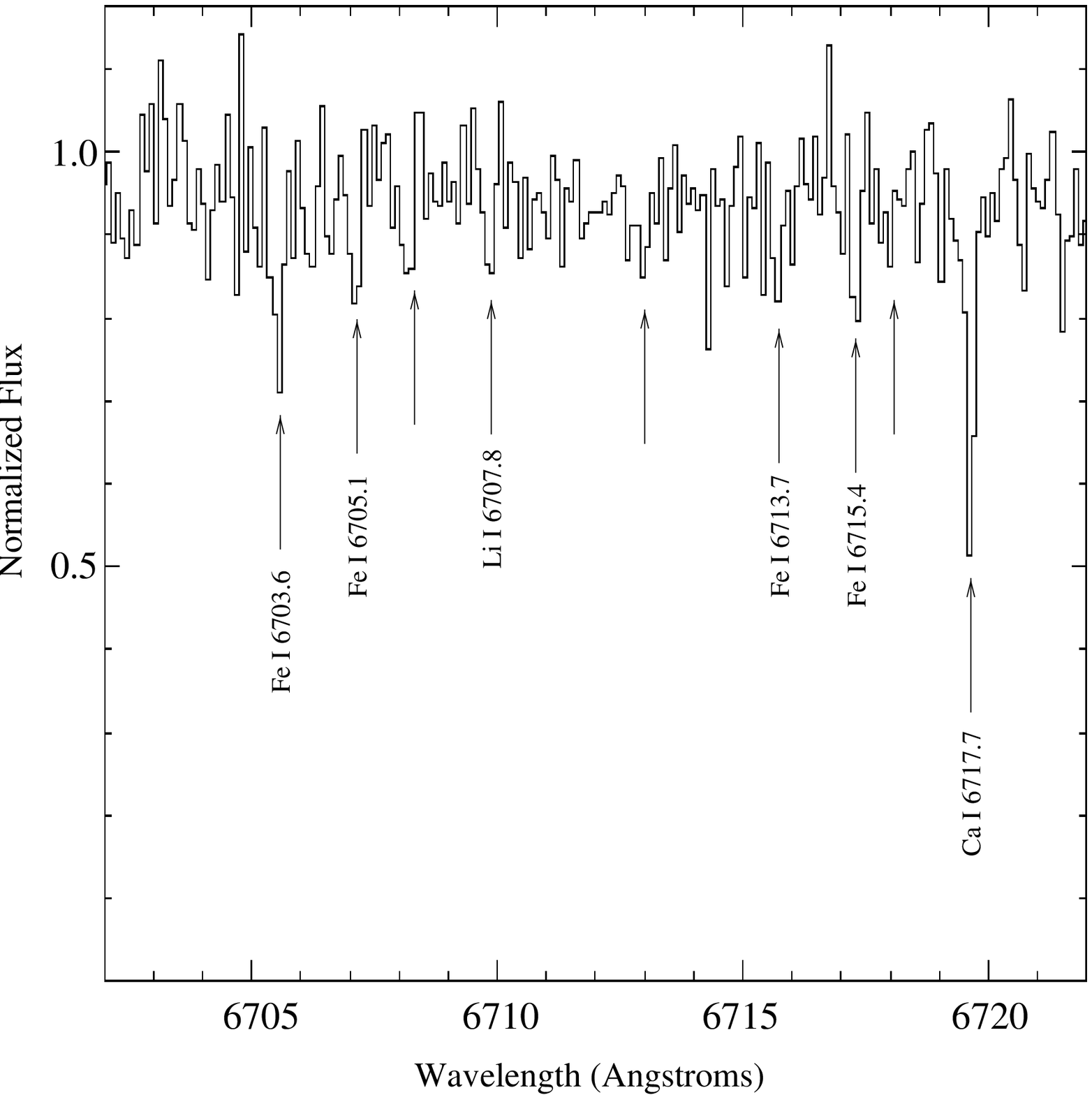}
\caption{
Portion of the spectrum with the lithium line. The lines of other elements
are also indicated for comparison (blends are not labelled). 
This spectrum has not been smoothed or binned.
}
\end{figure}

\begin{figure}
\plotone{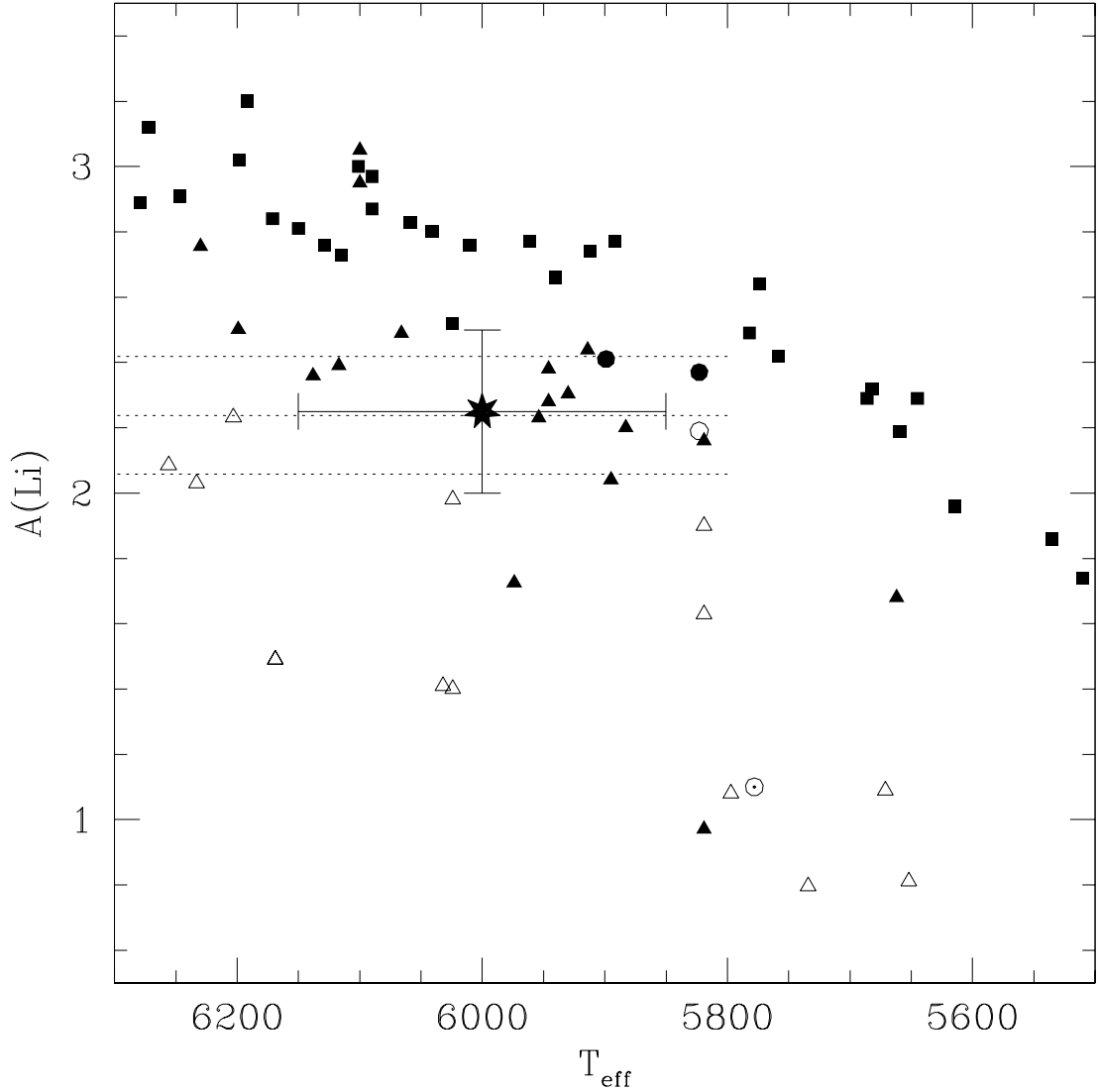}
\caption{ 
Lithium abundances $vs$ $T_{eff}$ for the Hyades (squares),
47Tuc (circles), M67 (triangles), 
the Sun ($\odot$), and 97BLG45 (big star). Full and open symbols indicate
detections and upper limits, respectively.
The primordial lithium abundance from Bonifacio \& Molaro (1997) is shown with
the middle dotted line, along with their $\pm 3 \sigma$ range given by the 
upper and lower lines.
}
\end{figure}

\begin{deluxetable}{rrrrrrrrrrrr}
\small
\footnotesize
\tablewidth{0pt}
\scriptsize
\tablecaption{Photometric Data on 97BLG45}
\tablehead{
\multicolumn{1}{c}{Star}&
\multicolumn{1}{c}{MACHO ID}&
\multicolumn{1}{c}{$RA_{2000}$}&
\multicolumn{1}{c}{$DEC_{2000}$}&
\multicolumn{1}{c}{$l$}&
\multicolumn{1}{c}{$b$}&
\multicolumn{1}{c}{$R_{gal}$}&
\multicolumn{1}{c}{$V$}&
\multicolumn{1}{c}{V$-$R}&
\multicolumn{1}{c}{$E_{V-R}$}&
\multicolumn{1}{c}{$M_V$}}
\startdata
97BLG45&142.27650.6057&18:21:01.8&--28:46:45&3.943&--6.680&1.08 kpc&19.80&0.60&0.20&+4.4\\
\enddata
\end {deluxetable}

\begin{deluxetable}{rrrrrrrrrrr}
\small
\footnotesize
\tablewidth{0pt}
\scriptsize
\tablecaption{Spectroscopic Data on 97BLG45}
\tablehead{
\multicolumn{1}{c}{Star}&
\multicolumn{1}{c}{Exptime}&
\multicolumn{1}{c}{Dispersion}&
\multicolumn{1}{c}{$S/N$}&
\multicolumn{1}{c}{$W_{Li 6707}$}&
\multicolumn{1}{c}{$W_{Fe 6703}$}&
\multicolumn{1}{c}{$W_{Fe 6705}$}&
\multicolumn{1}{c}{$W_{Fe 6713}$}&
\multicolumn{1}{c}{$A(Li)$}}
\startdata
97BLG45&6$\times$1800 sec&0.094 \AA/pix&50&$58\pm 15$m\AA &$94\pm 15$m\AA &$64\pm 15$m\AA &$68\pm 15$m\AA &$2.25\pm 0.25$\\
\enddata
\end {deluxetable}

\end{document}